\documentclass[journal]{IEEEtran}[letterpaper]

\usepackage[nolist]{acronym}
\usepackage{cite}
\usepackage{graphicx}
\usepackage{enumitem,kantlipsum}

\usepackage[left=0.63in,right=0.65in,top=0.7in,bottom=1.05in]{geometry}
\begin{document}
\begin{acronym}[]
    \acro{ML}{machine learning}
    \acro{UWB}{ultra-wideband}
    \acro{AI}{artificial intelligence}
    \acro{NLOS}{non-line-of-sight}
    \acro{WPFM}{wireless physical-layer foundation model}
    \acro{LLM}{large language model}
    \acro{3GPP}{3rd Generation Partnership Project}
    \acro{NLP}{neurolinguistic programming}
    \acro{IQ}{in-phase and quadrature}
    \acro{CIR}{channel impulse response}
    \acro{RSSI}{received signal strength indicator}
    \acro{SOTA}{state of the art}
\end{acronym}

\title{Towards a Wireless Physical-Layer Foundation Model: Challenges and Strategies}

\author{Jaron~Fontaine,
        Adnan~Shahid,~\IEEEmembership{Senior Member}
        and~Eli~De~Poorter%
\thanks{The authors are with IDLab, Ghent University.}%
}




\maketitle

\begin{abstract}
\Ac{AI} plays an important role in the dynamic landscape of wireless communications, solving challenges unattainable by traditional approaches. This paper discusses the evolution of wireless AI, emphasizing the transition from isolated task-specific models to more generalizable and adaptable AI models inspired by recent successes in \acp{LLM} and computer vision. To overcome task-specific AI strategies in wireless networks, we propose a unified \ac{WPFM}. Challenges include the design of effective pre-training tasks, support for embedding heterogeneous time series and human-understandable interaction. The paper presents a strategic framework, focusing on embedding wireless time series, self-supervised pre-training, and semantic representation learning. The proposed \ac{WPFM} aims to understand and describe diverse wireless signals, allowing human interactivity with wireless networks. The paper concludes by outlining next research steps for \acp{WPFM}, including the integration with \acp{LLM}.
\end{abstract}



\section{Introduction}

In the ever-evolving landscape of wireless communications and networking research, including the advancements of the upcoming 6G standard, \ac{AI} has become the focal point for tackling issues that cannot be solved by traditional non-machine learning approaches \cite{9446676}. Such approaches are based on traditional optimization theory techniques and only work if suitable mathematical models are available. With increasing user traffic and density, formulating such models is difficult \cite{zappone2019wireless}. Deep learning is suitable in such situations because it can find effective functions from a dataset, while aiming to be scalable and generalizable. It powers a wide range of applications such as recognizing wireless technologies, wireless optimization, classifying \ac{NLOS} conditions, estimating environmental parameters and analyzing human behavior in sports and healthcare applications. \cite{fontaine2019towards,cheng2021deep,fontaine2023transfer}. Recently, significant efforts have been made by \ac{3GPP} to standardize \ac{ML} in wireless networks. \ac{3GPP} has integrated \ac{ML} into the 5G core architecture in releases 15 and 16, while `Open RAN' embraces a \ac{ML}-native approach \cite{9768336}.

\begin{figure*}
    \centering
    \includegraphics[width=0.84\textwidth]{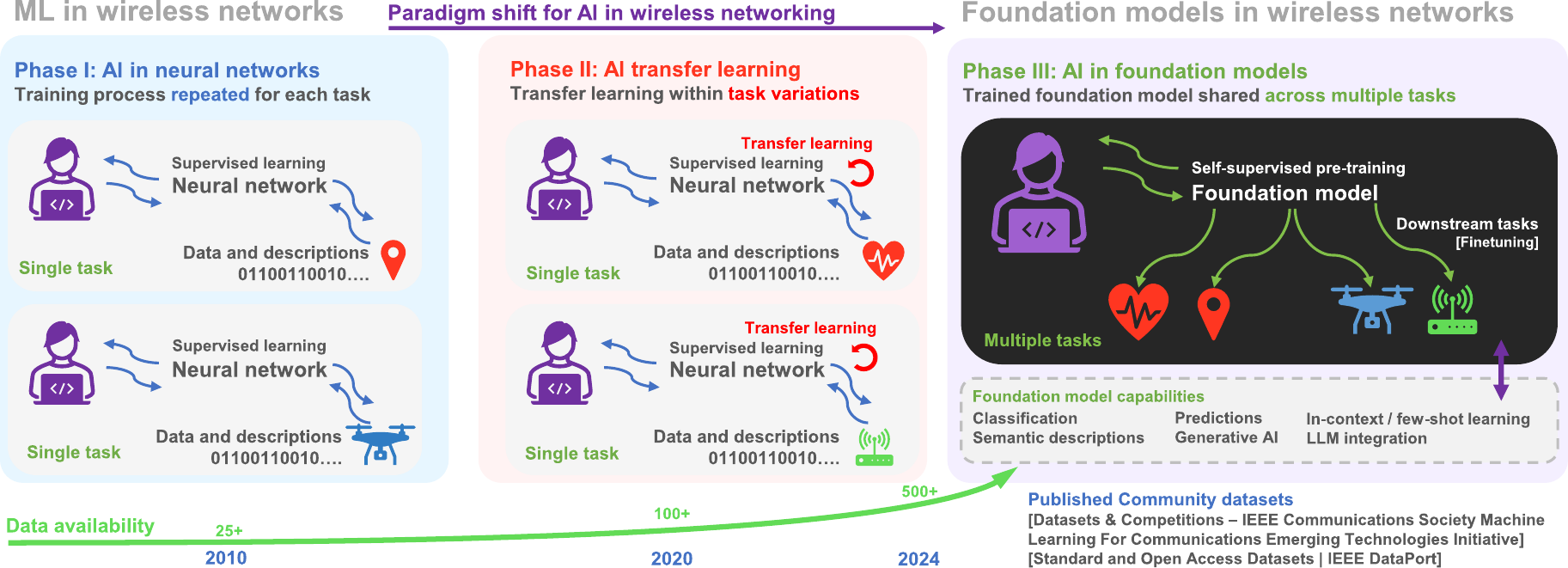}
    \caption{An overview of the three phases of deep learning, where a \ac{WPFM} aims to be a generalizable and adaptable model accross multiple wireless applications.}
    \label{fig:FigureOverview}
\end{figure*}

However, there is a fundamental issue with the predominant deep learning approach. A large number of research papers and solutions already exist, but are often developed and trained from scratch. Each of these thousands individual AI models operate in isolation which leads to a lack of knowledge sharing and reuse. If we look at the current trend of deep learning, which is presented in Figure \ref{fig:FigureOverview}, we can identify three phases and analyze how this issue evolved and can be addressed.


\begin{enumerate}[wide, labelwidth=!, labelindent=0pt]
    \item In the \textbf{first phase} of the development of wireless AI (see Figure 1), many researchers individually proposed \ac{ML} models for different wireless communication tasks such as recognition of wireless technologies, management of radio resources, classification of (N)LOS, etc., which require domain expertise and expensive \ac{ML} model training costs \cite{kulin2021survey}. Although these models use popular architectures such as deep, convolutional, and long short-term memory neural networks, the idea of sharing knowledge (weights) across various wireless downstream tasks is vastly underexplored.
    
    \item A \textbf{second phase} of advances in AI has been led by domains that have a lot of data available, such as \ac{NLP} and the computer vision domain. In such domains, neural networks have been thoroughly investigated and successfully demonstrated over the last decade. During this period, transfer learning has been proposed to leverage and synthesize the knowledge distilled within a task and use valuable experience accumulated in the past to facilitate the learning of new problems \cite{nguyen2021transfer}. This requires less labeled data, \ac{ML} can adapt quickly to new environments and can be more robust. However, this approach focuses on variations within single tasks, including those in the wireless domain. 
    
    \item Very recently, in the \textbf{third phase,} key AI domains have undergone a paradigm shift with the introduction of foundation models. A foundation model is a ``paradigm for building AI systems" in which a model trained on a large amount of unlabeled data can be adapted to many different downstream tasks, even with different types of data \cite{zhou2023comprehensive}. For example, a foundation model for image generation can cope with a wide variety of styles and shapes. Upon showing just a few new samples of a new style, the foundation model is adapted towards a new downstream task (e.g., generation of images using the new style) without significant effort required. 
    When we consider these rapid advances in AI, it becomes evident that current task-based ML approaches in wireless networks are no longer the way forward. 
\end{enumerate}

\vspace{-3pt}

As such, the evolution of AI development reveals a transition from task-specific models to the emergence of adaptable AI models that work across many tasks. 
To this end, this paper aims to contribute significantly by proposing foundation models for physical-layer wireless networks AI across multiple downstream tasks, thereby enabling future advancements in this domain. The main contributions presented in this paper are the following:

\begin{itemize}
    \item A brief overview of advances in foundation models and their applicability in the wireless domain.
    \item Analysis of missing steps and main challenges in the wireless domain to build a foundation model.
    \item A \acf{WPFM} framework.
    \item Strategies to apply wireless use cases effectively.
\end{itemize}

In the next section, we explore the advances in foundation models and discuss the need for \acp{WPFM} to streamline AI development in various tasks in wireless communication and compare its difference with \acp{LLM}. Next, Section \ref{sct:challenges} presents the challenges in the wireless domain of building a \ac{WPFM}. Section \ref{sct:validation} outlines wireless applications, used to demonstrate the foundation model in Section \ref{sct:architecture}, where we present a general \ac{WPFM} framework and implementation strategies. Next, Section \ref{sct:nextsteps} discusses the next research steps for \acp{WPFM} and enabling human-interactive wireless networks. Finally, conclusions are drawn in Section \ref{sct:conclusions}.

\section{Advances in foundation models }
\label{sct:advances}


\subsection{The rise of foundation models}
Foundation models have gained a lot of interest in the field of natural language processing and computer vision with the introduction of models such as GPT3/4, ChatGPT, BERT, DALL-E, etc. \cite{brown2020language}. These models have enabled significant progress and achieved \ac{SOTA} accuracy in their fields. With the abundant availability of large datasets, foundation models can be seamlessly applied to a diverse set of downstream tasks. Innovations to such foundation models include unsupervised pre-training techniques that enables models to grasp complex patterns and relationships within massive datasets. Additionally, fine-tuning strategies have improved in recent years, e.g. few- and zero-shot learning for new applications.

To accommodate diverse data types, multimodality has become a prominent feature in machine learning and is frequently adapted by foundation models \cite{liu2022zero}. Models such as `CLIP', designed for vision tasks, seamlessly merge text and image encodings, facilitating zero-shot image identification \cite{radford2021learning}. Additionally, text-to-image models like DALL-E demonstrate a profound synergy between language and vision models, incorporating encoded text prompts and image generation capabilities \cite{ramesh2021zero}. The autoregressive-trained transformer can simultaneously encode text and images, showcasing robust performance even in zero-shot evaluation scenarios. This integration of modalities not only enhances versatility, but also underscores the adaptability of (transformer-based) neural network architectures to handle varied input types effectively.

Why have these remarkable advances in AI not been applied in other domains? Unfortunately, these advances have thrived with the availability of large amounts of data, primarily drawn from the Internet, trained on large models. Acquiring such data quantities proves challenging in numerous domains, since the Internet does not harbor the necessary data as abundantly as it does textual and visual content.

One domain facing this challenge, i.e. robotics, has started to embrace the collection large dataset to achieve a general-purpose (foundation) model \cite{padalkar2023open}. Instead of relying on single-task AI, which only works well on a single robot involved in a single experiment, researchers have collaborated on achieving a general AI robot that has learned from a pool of many robots. Surprisingly, the authors found that multirobot data could be used with similar methods used in \acp{LLM}, provided that they follow the recipe of using large neural networks with large datasets. The dataset contains five hundred different skills and interactions with thousands of different objects, and allows training of a foundation model without requiring any special features for cross-embodiment. Remarkably, the global (foundation) model outperformed the robots trained in individual tasks. Finally, the authors have successfully integrated large vision and language models, which are finetuned to describe actions based on visual data from the robot. This allows prompt-based robotic actions, even with unseen objects.     

In addressing the challenges posed by the advent of 6G networks, encompassing Non-Terrestrial Networks (NTN), Terrestrial Networks (TN), mmWave/THz network, underwater networks, and advancements in Wireless Local Area Networks (WLANs) \cite{10.1145/3571072}, could a unified \ac{WPFM} serve as a viable solution to streamline and reduce redundancy in AI development across diverse wireless networks? Much like in the domain of robotics, similar challenges exist, where large datasets sourced from the Internet are scarce or too simplistic to represent complex wireless conditions. However, the availability of data is quickly catching up. Popular journals are seeking papers specific to datasets \cite{fischione2023data}, as depicted in Figure \ref{fig:FigureOverview}.


\subsection{Strategic differences between \acp{LLM} and \acp{WPFM}}
\acp{LLM} have already been popularized in wireless networks with examples of tasks that include answering healthcare-related questions, classifying \ac{3GPP} working group based on technical specifications, answering telecom-related questions, network configurations, etc. \cite{bariah2023large} \cite{tong2023ten}. However, since these models lack an understanding of physical wireless signals, we believe the development of a \ac{WPFM} is crucial and opens up a semantic understanding of the physical wireless environment.
%
%
Unlike text-based configurations in the higher layers of wireless networks, the realization of a \ac{WPFM} requires dealing with complex and dynamic data in the form of time series. The adoption of ML innovations can be significantly accelerated with a model that can understand and represent these time series. Trained foundation models using architectures such as transformer networks will eliminate the need for exhaustive research, domain expertise, and expensive data collection, as it enables improved representation learning across multiple downstream tasks.
Next, it will be possible to integrate a \ac{WPFM} with \acp{LLM} which focus on higher layers of the network stack. In these layers, pre-trained \acp{LLM}, such as ChatGPT, BERT, and Llama, have been very recently investigated and can already perform general tasks quite well \cite{zhou2023comprehensive} \cite{lin2023pushing} \cite{maatouk2023large}. 
System operators will no longer solely rely on traditional rule-based configurations programmed by engineers for individual tasks (e.g. defined channel selection and power allocation parameters for each application); they will be able to simply instruct (prompt) \acp{LLM} to configure networks or build applications using in-context learning (giving input and output examples) and utilize automatic optimization (without explicit programming) with physical-layer sensing \acp{WPFM}, as illustrated in Figure 1. 


\section{Challenges in wireless physical-layer foundation models}
\label{sct:challenges}
Recent scientific work has provided preliminary building blocks for \acp{WPFM}, such as transfer learning solutions and per-task data representations. However, since there is no general methodology to integrate these building blocks into an overall foundation model, significant research remains to be done. In addition, the usage of \acp{LLM} has been a recent research topic and even standardization within wireless communications, however, no physical-layer results have been presented and fundamental research challenges have yet to be identified \cite{lin2023pushing, maatouk2023large, bariah2023large, tong2023ten}. This paper addresses the following fundamental research challenges:

\subsubsection{Challenge 1: Novel methods are required for wireless pre-training tasks}
Modern ML architectures are capable of handling complex pattern learning and scaling of many parameters. The transformer neural network \cite{vaswani2017attention}, for example, excels in efficiently learning sequential data with parallel training. In other domains, foundation models exist which can perform general tasks, such as CLIP, DALL-E and \acp{LLM} such as GPT3, GPT4 and BERT, but also have been applied recently in time series \cite{wang2022transformer, wang2022wir}. However, in wireless networks, the transformer architecture is missing its potential by being trained to perform individual tasks. Instead, training large foundation transformer models requires effective un- and self-supervised pre-training tasks, e.g., in \acp{LLM} next-word predictions, masking, positional predictions, denoising, etc., are used \cite{zhou2023comprehensive}. This is key in enabling a shared foundation model for multiple downstream tasks (Figure 1). However, with limited data availability in the past and challenges in supporting heterogeneous data types, there is a gap in the wireless domain for effective pre-training tasks. 

\subsubsection{Challenge 2: \acp{WPFM} need to support the embedding of heterogeneous wireless time series with different lengths, sampling rates and data types}
Creating a foundation model for wireless time series, such as \ac{IQ} samples and \acp{CIR}, diverges significantly from models in the NLP (i.e. semantic understanding) and vision domains (i.e. 2D / 3D patterns) due to the heterogeneous temporal nature of the data. Unlike text or images, time-series data unfold over time, which presents challenges related to capturing embedding temporal dependencies, modeling variations, and addressing issues like fading channels and dynamic network conditions. Another contrast with \ac{SOTA} can be found in forecasting time series. Here, the focus is on predicting future trends based on historical data and seasonal trends, often seen in financial markets or climate studies, emphasizing precision and accuracy in predicting future outcomes. Analyzing wireless networks also proves to be much more difficult when dealing with heterogeneous wireless time series which are non-intuitive, in contrast to data in the aforementioned domains. Wireless \ac{SOTA} embedding is limited to reducing the dimensionality (compression) or mapping (in)complete data in a homogeneous space \cite{nguyen2021transfer}. Developing a \ac{WPFM} demands a deep understanding of time-sensitive data dynamics heterogeneous, i.e. supporting samples from use cases with different lengths, sampling rates and data types (\ac{IQ}, \ac{RSSI}, \ac{CIR}, metadata etc.), setting it apart as a compelling and distinctive domain within the broader AI landscape.

\subsubsection{Challenge 3: To cope with the nonintuitive nature of wireless time series data, \acp{WPFM} need human-understandable interaction and prompt-based optimization}
The semantic description of wireless networks and their environments has been a recent topic for standalone (wireless) machine learning models \cite{radford2021learning, li2019visualbert}. Combining textual descriptions with the first two challenges in a foundation model allows AI to be used in many downstream applications and opens up new possibilities. For example, by combining the embedded information from \acp{WPFM} with such semantic descriptions, we allow \acp{LLM} to autonomously understand wireless data from the physical layer, enable AI chain-of-thoughts and prompt-based applications and optimization. \ac{SOTA} lacks the ability for wireless optimization (\ac{LLM} + \ac{WPFM}) and is limited to \ac{LLM}-based configurations using standardization and specification documents \cite{kotaru2023adapting}.

\section{Applications of a wireless physical-layer foundation model}
\label{sct:validation}
To demonstrate the framework we propose in Section \ref{sct:architecture} and how it can address these challenges in wireless networks, we introduce two use cases where the AI improvements are predominately focused on wireless time series such as \ac{IQ} samples, CIR with different sample rates, input lengths, etc. 


\begin{enumerate}[wide, labelwidth=!, labelindent=0pt]
    \item \textbf{Activity recognition using UWB radar:} Activity recognition in healthcare leveraging \ac{UWB} radar can rely on \ac{CIR} data for precise movement detection. The use of ML in activity recognition applications, often including IMU sensors, facilitates automated real-time movement detection and improves healthcare monitoring. Including \ac{UWB} radar as a sensor can mitigate the need for persons to wear on-body sensors. Due to the complex nature of \ac{UWB} CIR signals, a high cost will be associated to develop one of many individual activity recognition applications. 
    \item \textbf{Wireless spectrum management:} Wireless spectrum management involves recognizing coexisting technologies within the same spectrum using \ac{IQ} samples, analysed by ML models. The increasing number of wireless technologies complicates AI-driven recognition. This becomes even more complicated by the diverse number of bandwidths at which the technologies are operating, which requires distinct sampling rates. Traditional training approaches for individual ML models become challenging and time-consuming due to these complexities. 
\end{enumerate}
As such, there is a need for a \ac{WPFM} that can understand features in wireless time series, including \ac{UWB} CIRs, \ac{IQ} samples, etc. This can alleviate costly engineering of individual models, simplifying ML in existing and new wireless applications.

\section{A wireless physical-layer foundation model strategic framework}
\label{sct:architecture}

\begin{figure}
    \centering
    \includegraphics[width=\columnwidth]{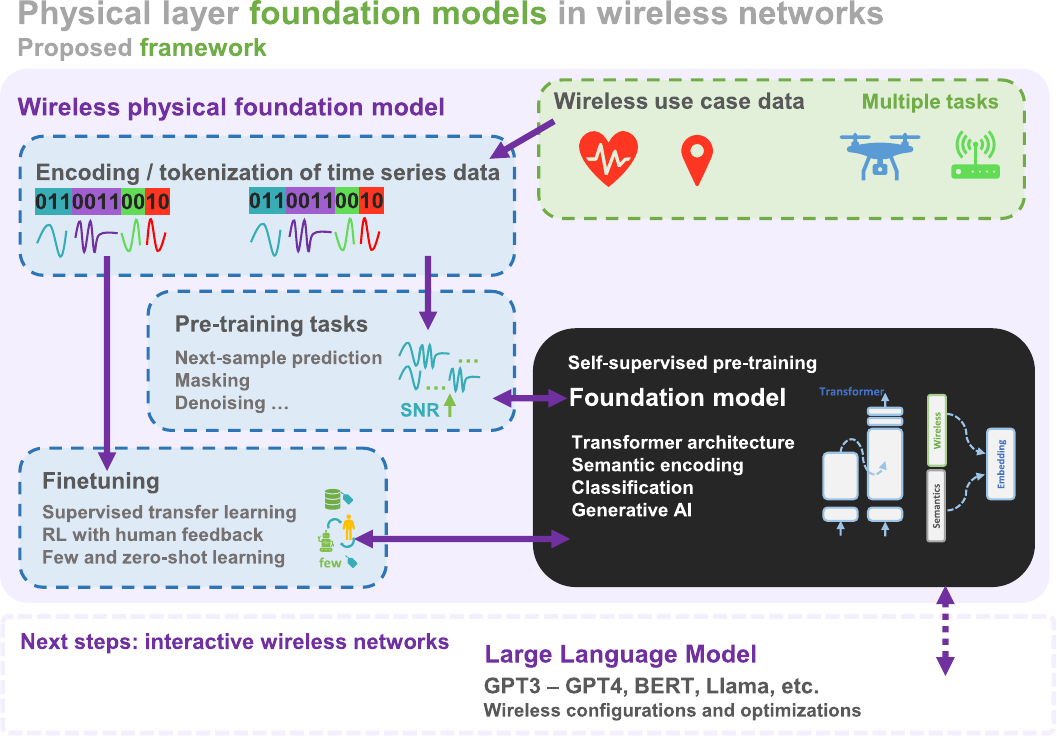}
    \caption{The different framework strategies of a \ac{WPFM}}
    \label{fig:architecture}
\end{figure}

To address the research challenges, we present a strategic framework of a \ac{WPFM}. This foundation model, illustrated in Figure \ref{fig:architecture} has a simple, but transformative goal: it can understand many types of physical wireless signals (time series) and metadata, while enabling network configurations and optimization and allowing user interaction. Using this framework and large openly available datasets, researchers can enable the usage of WPFM by companies and/or new researchers in the field to easily adapt WPFMs to new downstream tasks (e.g., \ac{CIR} understanding to detect different types of objects using \ac{UWB} radar, \ac{IQ} pattern encoding to recognize unseen wireless technologies in a spectrum manager and configuring a 6G network depending on the environmental conditions such as interference signals, noise, etc.).

\subsection{Embedding and tokenization of physical-layer time-series}
In the wireless domain, technologies and applications generally consist of heterogeneous time series. This means that the length of the input data, its data type and dimensionality can vary across applications. To ensure consistent input for foundation models, we propose using embedding or tokenizing of wireless time series data, which has several implementation options. One strategy is to use Byte-Pair Encoding (BPE) to merge frequently occurring signal or pulse \ac{IQ} sample pairs, until a fixed input size of the model is obtained. 
Another strategy is to represent known pulses and short-term patterns in time series as tokens. Alternatively, Time Variable based Tokenization (TVT) can be used to tokenize time series where the token uniformity inductive bias would act on the variable dimension \cite{zhou2022k}.
%
In addition to encoding, it is essential to define the vocabulary size (number of unique tokens) to support variable input lengths while avoiding unnecessary complexity. For instance, by applying tokenization techniques such as those utilized in models like BERT or GPT, we can ensure a uniform sample distribution regardless of the sampling rate. This results in a similar embedding space, e.g. $<$ 10\% difference, even when comparing signals sampled at different rates (e.g. 10 or 20 MHz) or when subjected to input noise with e.g. a signal-to-noise ratio (SNR) $>$ 10 dB. Such tokenization strategies are particularly valuable in scenarios involving signal processing at different bandwidths or sample rates and support data augmentation or simulation tasks. For example, they enable the generation of synthetic samples closely resembling wireless time series data, facilitating realistic simulations for various applications. The result is a robust tokenizer that can be integrated in well-defined foundation models such as transformers.

\textbf{Use case applications} 
In wireless use cases, these techniques allow CIRs with larger and shorter lengths to be understood by the framework, as well as technology recognition \ac{IQ} streams with different sampling rates and sample lengths. The shape of pulses or sine waves contains information about the wireless technology and modulation used or represents reflections in \ac{UWB} CIR data. These waves or pulses can be represented as tokens. To support longer or shorter time patterns in wireless applications, the accumulated token context size needs to be appropriate. Depending on the complexity of the transformer and its energy requirements, these can be extended or reduced.



\subsection{Effective self-supervised pre-training tasks for a \ac{WPFM}}
A crucial part of the development of foundation models is to obtain novel wireless pre-training tasks. %
%
Pre-training tasks are essential because they allow the model, e.g. a transformer, to learn general features and representations from a vast amount of diverse data before fine-tuning on a specific task. Allowing self-supervised pre-training is especially interesting in the wireless domain, which suffers from a low volume of labeled datasets. Pre-training can (i) facilitate learning general representation and contextual information that are applicable in multiple wireless applications, (ii) it can facilitate transfer learning to downstream tasks and (iii) it mitigates the need for large labeled datasets.
In wireless networks, we envision techniques such as (i) next sample/pulse/multipath prediction, aiming at predicting sequential information, (ii) masking which can learn to predict missing samples/pulse/multipath within the signal, (iii) denoising learns to remove (additive) Gaussian noise from signals, (iv) sequence order prediction learns to order signals in chronological order and (v) generative adversarial networks learn internal representations of the signal to generate realistic new versions of the signal under different conditions.
During training, several loss function strategies must be considered. One example is contrastive loss where, depending on the task, samples are assigned positive and negative labels. In the context of localization, samples collected near or at the same time and position can be considered positive, while others can be negative. Another approach is to assign positive labels to one sample with different augmentations and negative labels to other samples. 
%
By utilizing a pre-trained foundation model, we can effectively capture and extract common patterns within wireless time series data, enabling the application of generative AI techniques. Furthermore, it serves as a strong basis for further fine-tuning on specific wireless tasks.

\textbf{Use case applications}
Given an existing context of (embedded) CIRs or \ac{IQ} samples, the model can learn to predict the next likely pulse or sine wave samples, enhancing its ability to capture sequential information. Similarly, masking methods can teach the model to predict missing samples within the CIR signal or \ac{IQ} samples, improving its robustness to incomplete or noisy data. Finally, denoising can help train the model to remove Gaussian noise from signals, making it more resilient to real-world environmental conditions. When pre-training, contrastive loss can be assigned to CIR samples collected at the same time and position, to capture spatial and temporal correlations.

\subsection{Semantic physical-layer wireless representation learning}
To generate semantic information, we need to integrate textual descriptions of the wireless network and environment into the foundation model. These descriptions can be obtained from experts or pre-trained \acp{LLM}. To fuse text and wireless time series, multimodal fusion foundation models (transformers) can be designed. A simple approach is to share the latent space which enables joint representations of text and time series modalities. In contrast, separated latent space representations can be investigated which are more flexible and allow different architectures or model reuse for both text and time series. The textual description is different from the labels as it represents semantic information that can be provided along the wireless time-series data. For example, labels are the wireless technologies shared in the considered spectrum, while semantic textual descriptions can be the size and type of the environment, similarly to image captions in the computer vision domain. 
Learning semantic representations in wireless networks enables the foundation model to produce semantic information about the wireless network and environment in a textual form. In addition, it can also generate time series variations based on semantic descriptions in new environments and wireless technologies.

\textbf{Use case applications}
Metadata such as the dimensions of the environment (walls, ceiling height, etc.), type of environment (outdoor, office indoor, industrial environment, etc.), number of obstacles, persons, etc. can be included, as it impacts the efficiency and performance of wireless applications. Such descriptions can be augmented by \acp{LLM} and allow foundation models to generatively provide context to new, unknown environments. Additionally, generative capabilities can help with diagnostics or simulations of \ac{UWB} and wireless deployments.

\subsection{Fine-tuning \acp{WPFM}}
Although the foundation model may already be able to perform general tasks and applications, fine-tuning allows it to adapt the model to specific downstream tasks or domains. To enable this fine-tuning, we believe the following techniques are vital: (i) supervised transfer learning with a small, labeled dataset (tens of samples), (ii) reinforcement learning with human feedback (grading the predictions made by the foundation model before and during model deployment), (iii) zero- and few-shot learning by providing context (data context augmentation), and (iv) example input-output pairs (in-context learning) of the expected downstream task. Additionally, generative AI capabilities can enrich small datasets and allow these be used for fine tuning the model for the considered downstream tasks.

\textbf{Use case applications}
With reinforcement learning, a user of an activity recognition system can provide positive or negative feedback, while experts or network controllers can provide efficiency or performance feedback on wireless classification performance. In-context or zero- and few-shot learning can be integrated by giving in and output examples, e.g. a \ac{UWB} CIR + `falling down' movement, as part of a healthcare application. The model can take this example and explain new human activities detected by the \ac{UWB} radar.

\section{Next steps: the road to interactive wireless networks}
\label{sct:nextsteps}
The realization of a \ac{WPFM}, evaluated on a number of wireless applications, sparks several future research opportunities:

\begin{itemize}
    \item \textbf{LLM integration:} \acp{LLM} can be integrated with the \ac{WPFM}. This allows users to interact in a natural way with wireless networks. When users prompt statistics or the status of physical-layer conditions, the \ac{LLM} can request meta-data from the \ac{WPFM} which includes classifications, semantic descriptions of the wireless network conditions, sensing applications, human behavior, etc. This interaction requires further research to increase efficiency and raises open questions such as whether the pre-trained \ac{LLM} and \ac{WPFM} need to be finetuned together (training both latent spaces) or can coexist and interact on an API-basis.    
    \item \textbf{Multimodal \ac{WPFM}:} Next to supporting wireless time series and semantic meta-data of the environment, another modality that can be included is image-based information. As such, the \ac{WPFM} can incorporate and interact with images of floor plans, human activities, wireless access point or nodes topologies, etc.
    \item \textbf{Reasoning capabilities of WPFM and LLMs:} To understand why WPFMs and wireless LLMs make certain decisions or descriptions, it becomes crucial to implement reasoning capabilities (such as chain-of-thoughts), which is an active topic of investigation in the generative AI community \cite{sun2023survey}.  
    \item \textbf{Wireless network optimization:} With \ac{LLM} integration, the \ac{WPFM} enables continuous optimization of wireless networks, considering environmental and wireless conditions and providing this information to the \ac{LLM}. On the basis of this information, \acp{LLM} could output updated configurations and settings of the network. For example, when the \ac{WPFM} notices increased in human activity in a room, the \ac{LLM} can look at the \ac{UWB} configuration and update the sampling rate and bandwidth for increased accuracy. Such automated optimization can be enabled on a prompt basis, following any data sheets and standardization given as a context to existing \acp{LLM}.
    \item \textbf{Addressing energy consumption:} Foundation models tend to have a large number of parameters, resulting in very high complexity and energy consumption. At inference time, deploying such models in wireless networks is challenging. ML models benefit from reduced wireless bandwidth consumption when they are deployed close to the end device, but these are often resource-constrained \cite{woisetschlager2023federated}. From a training perspective, trade-offs should be investigated whether this requires more energy than combining multiple traditional task-focused ML solutions. By sharing the \ac{WPFM} and integrating in multiple domains and applications, the foundation model has the potential to save energy in the end.
    \item \textbf{Embedded and edge-based implementations:} Bringing foundation models closer to the end user is a trending topic within the research community. In massive (time critical) wireless networks, a large number of data samples are captured, each within their own setting and environment. Implementing federated learning at the edge or embedded devices enables rapid training and reduces the need to send massive amounts of data to a central repository \cite{10078088}. Bringing the knowledge from wireless federated learning into foundation models becomes crucial as the number of unique environmental conditions continues expand. 
\end{itemize}


\section{Conclusions}
\label{sct:conclusions}
This paper underscores the upcoming transformative work within the field of AI for wireless networks, evolving from task-specific models to the innovative paradigm of adaptable \acp{WPFM}. The challenges identified, such as effective pre-training tasks and embedding of heterogeneous time-series, highlight the complexities inherent in developing a unified model for wireless applications. The proposed \ac{WPFM} architecture aims to revolutionize AI development in wireless networks, accelerating and sharing AI advances. We believe the semantic capabilities of a \ac{WPFM} are a crucial step towards human interactive wireless networks. Such interactivity will be part of future research and can bridge the gap between expert-based manual configurations and automatic prompt-based configurations using \acp{LLM}. To realise the goal of a unified \ac{WPFM}, shared across many downstream tasks, upcoming research should focus on refining, standardizing pre-training tasks, tailored for the wireless domain. Additionally, real-world implementations are essential to empirically validate the effectiveness of the proposed \ac{WPFM}.




\ifCLASSOPTIONcaptionsoff
  \newpage
\fi



\bibliographystyle{ieeetr}
\bibliography{biblio.bib}

@article{zappone2019wireless,
  title={Wireless networks design in the era of deep learning: Model-based, AI-based, or both?},
  author={Zappone, Alessio and Di Renzo, Marco and Debbah, M{\'e}rouane},
  journal={IEEE Transactions on Communications},
  volume={67},
  number={10},
  pages={7331--7376},
  year={2019},
  publisher={IEEE}
}

@article{fontaine2019towards,
  title={Towards low-complexity wireless technology classification across multiple environments},
  author={Fontaine, Jaron and Fonseca, Erika and Shahid, Adnan and Kist, Maicon and DaSilva, Luiz A and Moerman, Ingrid and De Poorter, Eli},
  journal={Ad Hoc Networks},
  volume={91},
  pages={101881},
  year={2019},
  publisher={Elsevier}
}

@article{cheng2021deep,
  title={Deep learning for wireless networking: The next frontier},
  author={Cheng, Yu and Yin, Bo and Zhang, Shuai},
  journal={IEEE Wireless Communications},
  volume={28},
  number={6},
  pages={176--183},
  year={2021},
  publisher={IEEE}
}

@article{fontaine2023transfer,
  title={Transfer Learning for UWB error correction and (N) LOS classification in multiple environments},
  author={Fontaine, Jaron and Che, Fuhu and Shahid, Adnan and Van Herbruggen, Ben and Ahmed, Qasim Zeeshan and Abbas, Waqas Bin and De Poorter, Eli},
  journal={IEEE Internet of Things Journal},
  year={2023},
  publisher={IEEE}
}

@article{kulin2021survey,
  title={A survey on machine learning-based performance improvement of wireless networks: PHY, MAC and network layer},
  author={Kulin, Merima and Kazaz, Tarik and De Poorter, Eli and Moerman, Ingrid},
  journal={Electronics},
  volume={10},
  number={3},
  pages={318},
  year={2021},
  publisher={MDPI}
}

@ARTICLE{nguyen2021transfer,
  author={Nguyen, Cong T. and Van Huynh, Nguyen and Chu, Nam H. and Saputra, Yuris Mulya and Hoang, Dinh Thai and Nguyen, Diep N. and Pham, Quoc-Viet and Niyato, Dusit and Dutkiewicz, Eryk and Hwang, Won-Joo},
  journal={Proceedings of the IEEE}, 
  title={Transfer Learning for Wireless Networks: A Comprehensive Survey}, 
  year={2022},
  volume={110},
  number={8},
  pages={1073-1115},
  doi={10.1109/JPROC.2022.3175942}}

@article{zhou2023comprehensive,
  title={A comprehensive survey on pretrained foundation models: A history from bert to chatgpt},
  author={Zhou, Ce and Li, Qian and Li, Chen and Yu, Jun and Liu, Yixin and Wang, Guangjing and Zhang, Kai and Ji, Cheng and Yan, Qiben and He, Lifang and others},
  journal={arXiv preprint arXiv:2302.09419},
  year={2023}
}

@article{fischione2023data,
  title={Data Sets For Machine Learning In Wireless Communications And Networks},
  author={Fischione, Carlo and Chafii, Marwa and Deng, Yansha and Erol-Kantarci, Melike},
  journal={IEEE Communications Magazine},
  volume={61},
  number={9},
  pages={80--81},
  year={2023},
  publisher={IEEE}
}

@article{lin2023pushing,
  title={Pushing Large Language Models to the 6G Edge: Vision, Challenges, and Opportunities},
  author={Lin, Zheng and Qu, Guanqiao and Chen, Qiyuan and Chen, Xianhao and Chen, Zhe and Huang, Kaibin},
  journal={arXiv preprint arXiv:2309.16739},
  year={2023}
}

@article{maatouk2023large,
  title={Large language models for telecom: Forthcoming impact on the industry},
  author={Maatouk, Ali and Piovesan, Nicola and Ayed, Fadhel and De Domenico, Antonio and Debbah, Merouane},
  journal={arXiv preprint arXiv:2308.06013},
  year={2023}
}

@article{bariah2023large,
  title={Large Language Models for Telecom: The Next Big Thing?},
  author={Bariah, Lina and Zhao, Qiyang and Zou, Hang and Tian, Yu and Bader, Faouzi and Debbah, Merouane},
  journal={arXiv preprint arXiv:2306.10249},
  year={2023}
}

@article{tong2023ten,
  title={Ten issues of NetGPT},
  author={Tong, Wen and Peng, Chenghui and Yang, Tingting and Wang, Fei and Deng, Juan and Li, Rongpeng and Yang, Lu and Zhang, Honggang and Wang, Dong and Ai, Ming and others},
  journal={arXiv preprint arXiv:2311.13106},
  year={2023}
}

@article{vaswani2017attention,
  title={Attention is all you need},
  author={Vaswani, Ashish and Shazeer, Noam and Parmar, Niki and Uszkoreit, Jakob and Jones, Llion and Gomez, Aidan N and Kaiser, {\L}ukasz and Polosukhin, Illia},
  journal={Advances in neural information processing systems},
  volume={30},
  year={2017}
}

@article{wang2022transformer,
  title={Transformer-empowered 6G intelligent networks: From massive MIMO processing to semantic communication},
  author={Wang, Yang and Gao, Zhen and Zheng, Dezhi and Chen, Sheng and Gunduz, Deniz and Poor, H Vincent},
  journal={IEEE Wireless Communications},
  year={2022},
  publisher={IEEE}
}

@article{wang2022wir,
  title={WIR-Transformer: Using Transformers for Wireless Interference Recognition},
  author={Wang, Pengyu and Cheng, Yufan and Dong, Binhong and Hu, Ruofan and Li, Shaoqian},
  journal={IEEE Wireless Communications Letters},
  volume={11},
  number={12},
  pages={2472--2476},
  year={2022},
  publisher={IEEE}
}

@inproceedings{radford2021learning,
  title={Learning transferable visual models from natural language supervision},
  author={Radford, Alec and Kim, Jong Wook and Hallacy, Chris and Ramesh, Aditya and Goh, Gabriel and Agarwal, Sandhini and Sastry, Girish and Askell, Amanda and Mishkin, Pamela and Clark, Jack and others},
  booktitle={International conference on machine learning},
  pages={8748--8763},
  year={2021},
  organization={PMLR}
}

@article{li2019visualbert,
  title={Visualbert: A simple and performant baseline for vision and language},
  author={Li, Liunian Harold and Yatskar, Mark and Yin, Da and Hsieh, Cho-Jui and Chang, Kai-Wei},
  journal={arXiv preprint arXiv:1908.03557},
  year={2019}
}

@article{kotaru2023adapting,
  title={Adapting foundation models for information synthesis of wireless communication specifications},
  author={Kotaru, Manikanta},
  journal={arXiv preprint arXiv:2308.04033},
  year={2023}
}

@article{zhou2022k,
  title={A K-variate Time Series Is Worth K Words: Evolution of the Vanilla Transformer Architecture for Long-term Multivariate Time Series Forecasting},
  author={Zhou, Zanwei and Zhong, Ruizhe and Yang, Chen and Wang, Yan and Yang, Xiaokang and Shen, Wei},
  journal={arXiv preprint arXiv:2212.02789},
  year={2022}
}

@ARTICLE{9446676,
  author={Hoydis, Jakob and Aoudia, Fayçal Ait and Valcarce, Alvaro and Viswanathan, Harish},
  journal={IEEE Communications Magazine}, 
  title={Toward a 6G AI-Native Air Interface}, 
  year={2021},
  volume={59},
  number={5},
  pages={76-81},
  doi={10.1109/MCOM.001.2001187}}

@article{liu2022zero,
  title={Zero-shot learning via a specific rank-controlled semantic autoencoder},
  author={Liu, Yang and Gao, Xinbo and Han, Jungong and Liu, Li and Shao, Ling},
  journal={Pattern Recognition},
  volume={122},
  pages={108237},
  year={2022},
  publisher={Elsevier}
}

@inproceedings{ramesh2021zero,
  title={Zero-shot text-to-image generation},
  author={Ramesh, Aditya and Pavlov, Mikhail and Goh, Gabriel and Gray, Scott and Voss, Chelsea and Radford, Alec and Chen, Mark and Sutskever, Ilya},
  booktitle={International Conference on Machine Learning},
  pages={8821--8831},
  year={2021},
  organization={PMLR}
}

@article{brown2020language,
  title={Language models are few-shot learners},
  author={Brown, Tom and Mann, Benjamin and Ryder, Nick and Subbiah, Melanie and Kaplan, Jared D and Dhariwal, Prafulla and Neelakantan, Arvind and Shyam, Pranav and Sastry, Girish and Askell, Amanda and others},
  journal={Advances in neural information processing systems},
  volume={33},
  pages={1877--1901},
  year={2020}
}

@article{padalkar2023open,
  title={Open x-embodiment: Robotic learning datasets and rt-x models},
  author={Padalkar, Abhishek and Pooley, Acorn and Jain, Ajinkya and Bewley, Alex and Herzog, Alex and Irpan, Alex and Khazatsky, Alexander and Rai, Anant and Singh, Anikait and Brohan, Anthony and others},
  journal={arXiv preprint arXiv:2310.08864},
  year={2023}
}

@ARTICLE{9768336,
  author={Shehzad, Muhammad K. and Rose, Luca and Butt, M. Majid and Kovács, István Z. and Assaad, Mohamad and Guizani, Mohsen},
  journal={IEEE Vehicular Technology Magazine}, 
  title={Artificial Intelligence for 6G Networks: Technology Advancement and Standardization}, 
  year={2022},
  volume={17},
  number={3},
  pages={16-25},
  doi={10.1109/MVT.2022.3164758}}

@article{10.1145/3571072,
author = {Shen, Li-Hsiang and Feng, Kai-Ten and Hanzo, Lajos},
title = {Five Facets of 6G: Research Challenges and Opportunities},
year = {2023},
issue_date = {November 2023},
publisher = {Association for Computing Machinery},
address = {New York, NY, USA},
volume = {55},
number = {11},
issn = {0360-0300},
url = {https://doi.org/10.1145/3571072},
doi = {10.1145/3571072},
journal = {ACM Comput. Surv.},
month = {feb},
articleno = {235},
numpages = {39}
}

@article{sun2023survey,
  title={A survey of reasoning with foundation models},
  author={Sun, Jiankai and Zheng, Chuanyang and Xie, Enze and Liu, Zhengying and Chu, Ruihang and Qiu, Jianing and Xu, Jiaqi and Ding, Mingyu and Li, Hongyang and Geng, Mengzhe and others},
  journal={arXiv preprint arXiv:2312.11562},
  year={2023}
}

@article{woisetschlager2023federated,
  title={Federated fine-tuning of llms on the very edge: The good, the bad, the ugly},
  author={Woisetschl{\"a}ger, Herbert and Isenko, Alexander and Wang, Shiqiang and Mayer, Ruben and Jacobsen, Hans-Arno},
  journal={arXiv preprint arXiv:2310.03150},
  year={2023}
}

@ARTICLE{10078088,
  author={Pei, Jiaming and Li, Shike and Yu, Zhi and Ho, Laishan and Liu, Wenxuan and Wang, Lukun},
  journal={IEEE Communications Standards Magazine}, 
  title={Federated Learning Encounters 6G Wireless Communication in the Scenario of Internet of Things}, 
  year={2023},
  volume={7},
  number={1},
  pages={94-100},
  doi={10.1109/MCOMSTD.0005.2200044}}

\end{document}